\documentclass[reprint, showpacs, amsmath, amssymb, aps]{revtex4-1}
\usepackage{mathrsfs}
\usepackage{amsfonts}
\usepackage{amsmath,amsfonts,amssymb,mathrsfs,bm}
\usepackage{verbatim,psfrag,times,CJK}
\usepackage{graphicx,graphics,color,epsfig}
\usepackage{float}
\usepackage{flafter}
\usepackage{subeqnarray}
\usepackage{cases}

\begin{document}
\title{Squeezing of a movable mirror via the dissipative optomechanical coupling}
\author{Wen-ju Gu}
\affiliation{Department of Physics, Huazhong Normal University, Wuhan 430079, China.}
\author{Gao-xiang Li}
\email[]{gaox@phy.ccnu.edu.cn}
\affiliation{Department of Physics, Huazhong Normal University, Wuhan 430079, China.}
\author{Ya-ping Yang}
\affiliation{Department of Physics, Tongji University, Shanghai 200092, China.}

\begin{abstract}
We investigate the squeezing for a movable mirror in the dissipative optomechanics in which the oscillating mirror modulates both the resonance frequency and the linewidth of the cavity mode. Via feeding a much weaker broadband squeezed vacuum light accompanying the coherent driving laser field into the cavity, the master equation for the cavity-mirror system is derived by following the general reservoir theory based on the density operator in which the reservoir variables are adiabatically eliminated by using the reduced density operator for the system in the interaction picture. When the mirror is weakly coupled to the cavity mode, we find that under the conditions of laser cooling to the ground state, the driven cavity field can effectively perform as a squeezed vacuum reservoir for the movable mirror via utilizing the completely destructive interference of quantum noise, and thus the efficient transfer of squeezing from the light to the movable mirror occurs, which is irrespective of the ratio between the cavity damping rate and the mechanical frequency. When the mirror is moderately coupled to the cavity mode, the photonic excitation can preclude the completely destructive interference of quantum noise, and as a consequence, the mirror deviates from the ideal squeezed state.
\end{abstract}


\maketitle
\section{Introduction}
Rapid progress on optomechanics towards sensing and control of the zero-point motion of mechanical oscillators has been made via the engineering of high-quality micromechanical oscillators coupled to high-finesse cavity modes~\cite{I.WalsonRae, F.Marquardt1, T.J.Kippenberg, F.Marquardt2}, because exploration of quantum behavior in these mechanical systems will spark new insights into quantum information processing~\cite{V.Fiore,K.Stannigel,haoxiong}, measurement science~\cite{P.Verlot,E.Gavartin,JianQiZhang}, and fundamental tests of physical laws~\cite{I.Pikovski}, etc. Recently, some experimental investigations for observing quantum mechanical effects in the mechanical systems have been demonstrated~\cite{A.Schliesser, L.Mazzola, DWC.Brooks}. Indeed, these technical developments also open the possibility to observe nonclassical state of the mechanical oscillator~\cite{S.Rips}. Specifically, achieving squeezed states in mechanical oscillators, in which the variance of one quadrature of motion is below the zero-point motion, is an important goal because of its applications in ultrahigh precision measurements such as the detection of gravitational waves~\cite{C.M.Caves, A.Abramovici, B.C.Barish}. By now, different schemes have been proposed for the generation of quantum squeezing of movable mirrors~\cite{K.Jahne,A.Mari,HuatangTan,J.Q.Liao,Sumei,H.Seok}. For example, Huang~\emph{et al.}~\cite{Sumei} proposed a potential scheme to generate squeezing by putting an optical parameter amplifier inside a cavity, Seok~\emph{et al.}~\cite{H.Seok} presented a theoretical analysis of the motional squeezing of a cantilever magnetically coupled to a classical tuning fork via microscopic magnetic dipoles, and J\"ahne~\emph{et al.}~\cite{K.Jahne} investigated the creation of squeezed states of movable mirror transferred from a squeezed light driving the cavity via the dispersive coupling under the assumption of the resolved-sideband limit.

However, from a practical perspective, it is preferable to deviate from the resolved-sideband limit, since it allows one to use small drive detunings compared with the cavity decay rate and achieve much larger effective cavity-mechanical oscillator couplings. Recently, the dissipative cavity-mirror systems in both microwave and optical domains have been investigated, in which the driven cavity can effectively act like a zero-temperature bath via a destructive interference of quantum noise in the non-sideband-resolved regime and hence the ground-state cooling and low-power quantum-limited position transduction are both possible~\cite{F.Elste, A.Xuereb}. In addition, the enhanced cooling rate and elimination of optically-induced heating will be benefit for squeezing transfer from the squeezed light driving the cavity, as mechanical squeezing is fairly vulnerable to thermal and optically-induced heating scattering mechanisms. Thus, in this paper we will show that dissipative optomechanics can improve the performance of the squeezing transfer under the condition of the perfect elimination of heating process and finally lead to better mechanical squeezing.

In this paper, we propose a scheme that is capable of generating mechanical squeezing via engineering reservoir in an optomechanical setup having a strong dissipative coupling, which consists of an effective Fabry-P\'erot interference (FPI) with one movable ideal end mirror. The equivalent FPI is derived from a Michelson-Sagnac interference (MSI) with a movable membrane, explicitly shown in Refs.~\cite{K.Yamamoto,D.Friedrich,A.Xuereb}. When we feed a much weaker broadband squeezed vacuum light accompanying the coherent driving cooling-laser field into the cavity, the cavity field couples to the movable mirror via both the tunable dispersive and dissipative interactions. Then, distinct from the common Heisenberg-Langevin approach adopted in Refs.~\cite{A.Xuereb,F.Elste,K.Jahne,Sumei}, we follow the general reservoir theory based on the density operator in which the reservoir variables are adiabatically eliminated by using the reduced density operator for the system in the interaction picture. When the movable mirror is weakly coupled to cavity mode, the master equation for the movable mirror is derived by adiabatically eliminating the cavity field. It is shown that under the conditions of laser cooling to the ground motional state as discussed in Refs.~\cite{A.Xuereb,F.Elste}, i.e. elimination of the heating scattering process due to the completely destructive interference of quantum noise, the driven cavity can effectively perform as a squeezed vacuum reservoir for the movable mirror, and thus the efficient transfer of squeezing from the light squeezing to the movable mirror occurs, which is irrespective of the ratio between the cavity damping rate and the mechanical frequency. In addition, when the mirror is moderately coupled to the cavity mode, we solve the full motional equations for cavity-mirror system with a purely dissipative optomechanical coupling, we find that the photonic excitation can preclude the completely destructive interference of quantum noise and the mirror deviates from the ideal squeezed state. However, this dissipative optomechanics is still effective in squeezing the movable mirror around its ground mechanical state beyond weak-coupling regime as numerically shown. 

The paper is structured as follows. In Sec.II we introduce the FPI and derive the motion equation for the mirror-cavity system via adiabatically eliminating the reservoir variables. In Sec.III we analyze the cooling and squeezing of the movable mirror in the weak-coupling regime and results beyond the weak-coupling regime are presented in Sec.IV. In the last the conclusion is drawn in Sec.V.

\section{dissipative optomechanical system driven by a squeezed reservoir}
\subsection{Description of the dissipative optomechanical system}
We consider an optomechanical system consisted of an effective Fabry-P\'erot interferometer (FPI), sketched in Fig.1, which can be formed from the Michelson-Sagnac interferometer (MSI) with a movable membrane~\cite{K.Yamamoto,D.Friedrich,A.Xuereb}. The movable mirror $\mathcal{M}$ oscillates along the $x$-axis with the frequency $\omega_m$ and couples to a cavity mode with the resonant frequency $\omega_a$ via the dispersive and dissipative couplings, which corresponds to the shifts of the cavity's resonant frequency and damping rate respectively due to the mechanical motion. The full Hamiltonian is a sum of the free cavity $H_\textrm{c}$, free movable mirror $H_\textrm{m}$, free reservoir field $H_\textrm{R}$, cavity-reservoir interaction $H_\textrm{c-R}$ and cavity-mirror interaction $H_\textrm{int}$ Hamiltonians ($\hbar=1$):
\begin{subequations}
\begin{align}
&H=H_\textrm{c}+H_\textrm{m}+H_\textrm{R}+H_\textrm{c-R}+H_\textrm{int},\\
&H_\textrm{c}=\omega_a a^\dag a,\\
&H_\textrm{m}=\omega_m b^\dag b,\\
&H_\textrm{R}=\int d\omega \omega a^\dag_\omega a_\omega,\\
&H_\textrm{c-R}=i\sqrt{\frac{\kappa_c}{\pi}}\int d\omega(a^\dag_\omega a-a^\dag a_\omega),\\
&H_\textrm{int}=g_0\bigg[\alpha a^\dag a+i\beta\sqrt{\frac{L}{2\pi c}}\int d\omega(a^\dag_\omega a-a^\dag a_\omega)\bigg](b+b^\dag).
\end{align}
\end{subequations}
The operators $a$ and $b$ are the annihilation operators of cavity and phonon modes. The operator $a_\omega$ describes the continuous modes of optical reservoir coupled to the cavity mode and $\kappa_c$ is the damping rate of the cavity field without the motion of the mirror. The parameters $\alpha$ (dispersive) and $\beta$ (dissipative) respectively represent the cavity frequency's ($\omega_a$) and damping rate's ($\kappa_c$) linear dependence on the small displacement $x$ with $x=x_0(b^\dag+b)/\sqrt{2}$, where $x_0$ is the zero-point motion amplitude of the movable mirror. The effective length of the interferometer is $L$. This optomechanical setup can realize the strong dissipative coupling, even in the order of cavity linewidth in the absence of dispersive coupling, i.e. $\alpha=0$~\cite{A.Xuereb}.
\begin{figure}[hbt]
\centering\includegraphics[width=6cm,keepaspectratio,clip]{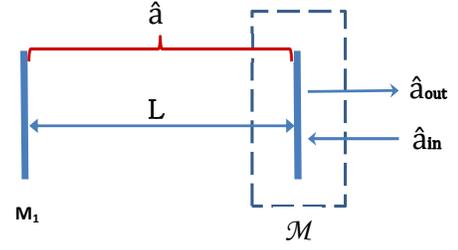}
\caption{ (Color online) Sketch of the effective Fabry-P\'erot interferometer (FPI) coupled to the cavity mode via the dispersive and dissipative couplings. The cavity is driven by a weaker squeezed vacuum field accompanying a coherent driving laser.}
\end{figure}

The dispersively and dissipatively coupled optomechanical system has been investigated to cool the mechanical oscillator to its ground state in microwave and optical domains in the Heisenberg-Langevin approach~\cite{A.Xuereb,F.Elste,T.Weiss}. In this paper, we present the dynamics of the movable mirror based on the density operator in which the reservoir and cavity variables can be adiabatically eliminated by using the reduced density operator for the system. The optical reservoir has two contributions on the cavity field: the c-number part $\langle a_\omega\rangle=\sqrt{2\pi}\bar{a}_\textrm{in}e^{-i\omega_Rt}$ corresponding to coherent cooling laser of frequency $\omega_R$ and random noise part $\delta a_{\omega}$ describing the broadband squeezed vacuum reservoir with central frequency $\omega_s$. The noise operator $\delta a_\omega$ has zero mean value and second moments are~\cite{C.W.Gardiner}
\begin{align}
&\langle \delta a^\dag_\omega \delta a_{\omega'}\rangle=N\delta(\omega-{\omega'}),\nonumber\\
&\langle \delta a_\omega \delta a^\dag_{\omega'}\rangle=(N+1)\delta(\omega-{\omega'}),\nonumber\\
&\langle \delta a_\omega \delta a_{\omega'}\rangle=M\delta(\omega+{\omega'}-2\omega_s),\nonumber\\
&\langle \delta a^\dag_\omega \delta a^\dag_{\omega'}\rangle=M^*\delta(\omega+{\omega'}-2\omega_s),
\end{align}
where $N=\sinh^2(r)$, $M=\sinh(r)\cosh(r)e^{i\varphi}$ with $r$ the squeezing parameter of the squeezed vacuum light and $\varphi$ the phase of the squeezed vacuum light.

\subsection{Adiabatically elimination of the squeezed reservoir}
Since a broadband squeezed vacuum is assumed, where the bandwith of the squeezed reservoir is not only larger than typical spontaneous dissipation rates of the cavity field but also large compared to detunings and the effective coupling strength between the cavity and mirror, the Markovian master equation for the cavity-mirror system is obtained via adiabatically eliminating the squeezed vacuum reservoir variables~\cite{H.J.Carmicael}. Following the general reservoir theory in textbook~\cite{M.O.Scully}, the system-reservoir interaction is given by $\mathscr{V}(t)=H_\textrm{c-R}(t)+H_\textrm{int}(t)$ in the interaction picture. By tracing over the reservoir coordinates, the reduced density operator $\rho_s$ for the cavity-mirror system is governed by the equation
\begin{align}
\dot{\rho}_s=&-i\textrm{Tr}_R[\mathscr{V}(t), \rho_s(t)\otimes\rho_R(t_i)]\nonumber\\
&-\textrm{Tr}_R\int_{t_i}^t[\mathscr{V}(t),[\mathscr{V}(t'),\rho_s(t)\otimes\rho_R(t_i)]]dt'.
\end{align}

By substituting the c-number component and the two-time correlation functions in Eq.(2) into the Eq.(3), the motion equation for the density operator $\rho_s$ can now be obtained as
\begin{align}
\dot{\rho}_s=-i[H_0,\rho_s]+\mathcal{L}_1\rho_s+\mathcal{L}_2\rho_s,
\end{align}
where the Hamiltonian $H_0$ consists of the free Hamiltonians of cavity and phonon modes, which is given by
\begin{align}
H_0=-\Delta a^\dag a+\omega_m b^\dag b
\end{align}
with $\Delta=\omega_R-\omega_a$ the detuning of the cavity resonant frequency from the coherent driving light frequency, and the Liouvillian operators $\mathcal{L}_2$ and $\mathcal{L}_1$ include the dissipations of the cavity and phonon modes and interactions between them, which are expressed as
\begin{align}
\mathcal{L}_1\rho_s&=-i[g_0\alpha a^\dag a(b+b^\dag)+i\sqrt{2}(\bar{a}^*_{in}C-\bar{a}_{in}C^\dag),\rho_s],\nonumber\\
\mathcal{L}_2\rho_s&=M^*e^{i2\Delta_st}(C^2\rho_s+\rho_sC^2-2C\rho_sC)\nonumber\\
&+Me^{-i2\Delta_st}({C^\dag}^2\rho_s+\rho_s{C^\dag}^2-2C^\dag\rho_sC^\dag)\nonumber\\
&+N(2C^\dag\rho_sC-CC^\dag\rho_s-\rho_sCC^\dag)\nonumber\\
&+(N+1)(2C\rho_sC^\dag-C^\dag C\rho_s-\rho_sC^\dag C),
\end{align}
with the composite operator $C=[\sqrt{\kappa_c}+g_0\beta\sqrt{\frac{L}{2c}}(b+b^\dag)]a$ and $\Delta_s=\omega_s-\omega_R$ the detuning between the central frequency of squeezing vacuum reservoir and the coherent driving light frequency.

The master equation in Eq.(4) is difficult to be exactly solved because of the existence of the nonlinear terms. However, outside the strong-coupling regime as discussed in single-photon optomechanics~\cite{A.Nunnenkamp}, it is valid to proceed the linearization on the full quantum dynamics by assuming that each operator in the system can be written as the sum of its mean value and a small fluctuation~\cite{W.J.Gu}:
\begin{align}
a=\bar{a}+d, \hspace{10pt}b=\bar{b}+f,
\end{align}
where the classical components $\bar{a}=\langle a\rangle$, $\bar{b}=\langle b\rangle$. In the paper, our considerations are explicitly focused outside the strong-coupling limit, i.e. the optomechanical coupling strengths $(g_0\alpha, g_0\beta\sqrt{\kappa_cL/2 c})\ll(\omega_m,\kappa_c)$. Thus to the lowest order of the strengths $g_0\alpha$ and $g_0\beta\sqrt{\kappa_cL/2 c}$, the mean phonon operator $\bar{b}\approx0$ and the mean cavity operator $\bar{a}$ obeys the equation
\begin{align}
\frac{d}{dt}\bar{a}=(i\Delta-\kappa_c)\bar{a}-\sqrt{2\kappa_c}\bar{a}_{in}.
\end{align}
The steady-state solution for $\bar{a}$ is obtained as
\begin{align}
\bar{a}=\frac{\sqrt{2\kappa_c}}{i\Delta-\kappa_c}\bar{a}_{in}.
\end{align}

In this shifted representation, the evolution of the cavity-mirror system is governed by the contributions respectively induced by the motions for the uncoupled cavity and phonon modes and the interaction between them, which is given by
\begin{align}
\frac{d}{dt}\rho_s=\mathcal{L}^\textrm{d}\rho_s+\mathcal{L}^\textrm{f}\rho_s+\mathcal{L}^\textrm{d-f}\rho_s.
\end{align}
The uncoupled cavity contribution obeys the equation
\begin{align}
\mathcal{L}^\textrm{d}\rho_s&=i[\Delta d^\dag d,\rho_s]+\kappa_cM^*e^{i2\Delta_st}(d^2\rho_s+\rho_sd^2-2d\rho_sd)\nonumber\\
&+\kappa_cMe^{-i2\Delta_st}({d^\dag}^2\rho_s+\rho_s{d^\dag}^2-2{d^\dag}\rho_s{d^\dag})\nonumber\\
&+\kappa_cN(2d^\dag\rho_sd-dd^\dag\rho_s-\rho_sdd^\dag)\nonumber\\
&+\kappa_c(N+1)(2d\rho_sd^\dag-d^\dag d\rho_s-\rho_sd^\dag d),
\end{align}
which parallels the evolution of the field in a cavity coupled to an outside squeezed vacuum reservoir. The uncoupled mirror follows the equation
\begin{widetext}
\begin{align}
\mathcal{L}^\textrm{f}\rho_s&=-i[\omega_mf^\dag f,\rho_s]
+g_0^2\beta^2\frac{L}{2c}\big[(2N+1)\vert\bar{a}\vert^2-M^*e^{i2\Delta_st}\bar{a}^2-Me^{-i2\Delta_st}\bar{a}^{\ast2}\big]\nonumber\\&
\times\big[2(f+f^\dag)\rho_s(f+f^\dag)-(f+f^\dag)^2\rho_s-\rho_s(f+f^\dag)^2\big].
\end{align}
The interaction between the cavity field and movable mirror is described by
\begin{align}
\mathcal{L}^\textrm{d-f}\rho_s&=-i g_0\big[\Big(\alpha(\bar{a}^*d+\bar{a}d^\dag)+i\beta\sqrt{\frac{L}{c}}(\bar{a}^*_{in}d-\bar{a}_{in}d^\dag)\Big)(f+f^\dag),\rho_s\big]\nonumber\\
&+2\Big\{g_\textrm{eff}M^*e^{i2\Delta_st}\big[d(f+f^\dag)\rho_s+\rho_sd(f+f^\dag)-(f+f^\dag)\rho_sd-d\rho_s(f+f^\dag)\big]+h.c.\Big\}\nonumber\\
&+2\Big\{g_\textrm{eff}N\big[d^\dag\rho_s(f+f^\dag)-\rho_sd^\dag(f+f^\dag)\big]+g_\textrm{eff}(N+1)\big[(f+f^\dag)\rho_sd^\dag-d^\dag(f+f^\dag)\rho_s\big]+h.c.\Big\},
\end{align}
\end{widetext}
with
\begin{align}
g_\textrm{eff}=g_0\beta\sqrt{\frac{\kappa_cL}{2c}}\bar{a}
\end{align}
the effective dissipative coupling strength between the cavity field and movable mirror. Obviously, the effective dispersive coupling strength is characterized by $g_0\alpha\bar{a}$.
\section{cooling and squeezing for the movable mirror in the weakly coupling regime}
\subsection{Adiabatically elimination of the cavity field}
In the weakly optomechanical coupling regime, in which the cavity field weakly couples to the movable mirror such that the effective strengths $g_0\alpha\bar{a}$ and $g_\textrm{eff}$ are much smaller than cavity damping rate $\kappa_c$, the cavity variable arrives at the steady state much faster than the mirror variable and can be adiabatically eliminated. Thus, the equation of motion for the deduced density operator of the movable mirror can be also calculated  paralleling the method for derivation of the cavity-mirror system in the last subsection by tracing over the cavity variable. Applying the second-order perturbation method with respect to the effective coupling strengths $g_0\alpha\bar{a}$ and $g_\textrm{eff}$, the reduced density operator for the movable mirror $\rho_f$ now becomes
\begin{align}
\frac{d}{dt}\rho_f=\textrm{Tr}_\textrm{d}\int^t_{t_0}\mathcal{L}^\textrm{d-f}(t)\mathcal{L}^\textrm{d-f}(t')\rho_d(t_0)\otimes\rho_f(t)dt',
\end{align}
where $\rho_d(t_0)$ is the steady-state density operator of cavity field, governed by the Liouvillian operator in Eq.(11). With the definition of the detuning $\delta=\Delta_s-\omega_m$ and assumption of $\delta\ll(\Delta_s,\omega_m)$ to accommodate for cavity induced energy shift, after some calculations the resulting equation of motion for the mirror described by the reduced density matrix $\rho_f$ is
\begin{align}
&\frac{d}{dt}\rho_f=-i[H_f,\rho_f]\nonumber\\&+\Big[\Theta(\omega_m)M^*e^{i2\delta t}(2f\rho_ff-f^2\rho_f-\rho_ff^2)+h.c.\Big]\nonumber\\&+\Big[N\vert\Theta(\omega_m)\vert+(N+1)\vert\Theta(-\omega_m)\vert\Big]
(f^\dag\rho_ff-ff^\dag\rho_f+h.c.)\nonumber\\&+\Big[N\vert\Theta(-\omega_m)\vert+(N+1)\vert\Theta(\omega_m)\vert\Big]
(f\rho_ff^\dag-f^\dag f\rho_f+h.c.)
\end{align}
in the rotating-wave approximation, with
\begin{align}
\Theta(\omega_m)=\frac{g_\textrm{eff}^2}{\kappa_c}\frac{(2\Delta+\omega_m+\frac{\alpha}{\beta}\sqrt{\frac{2\kappa_cc}{L}})^2}{[i(\Delta+\omega_m)+\kappa_c]^2},
\end{align}
where the optically-induced energy shift for the phonon frequency is indicated by the Hamiltonian $H_f$
\begin{align}
H_f&=\frac{\vert g_\textrm{eff}\vert^2}{\kappa^2_c}\bigg\{\Big[(\frac{\alpha}{\beta}\sqrt{\frac{2\kappa_cc}{L}}+\Delta)^2+\kappa_c^2\Big]\Big[\theta_1(-\omega_m)
+\theta_1(\omega_m)\Big]\nonumber\\&-2\kappa_c^2\Big[\theta_2(\omega_m)+\theta_2(-\omega_m)\Big]\bigg\}f^\dag f,
\end{align}
with $\theta_1(\omega_m)=(\Delta+\omega_m)/[(\Delta+\omega_m)^2+\kappa_c^2]$, $\theta_2(\omega_m)=(2\Delta+\omega_m+\frac{\alpha}{\beta}\sqrt{\frac{2\kappa_cc}{L}})/[(\Delta+\omega_m)^2+\kappa_c^2]$. In general, when $\Theta(-\omega_m)=0$, i.e. the detuning fulfills the relation
\begin{align}
\Delta=\omega_m/2-\frac{\alpha}{\beta}\sqrt{\frac{2\kappa_cc}{L}}/2,
\end{align}
which is just the optimal detuning for ground-state cooling of mechanical oscillator appeared in the dissipative optomechanics and simultaneously the detuning $\delta$ fulfills the relation
\begin{align}
\delta=\frac{\vert g_\textrm{eff}\vert^2}{\kappa^2_c}\frac{2\Delta(\Delta^2-\omega_m^2+\kappa_c^2)-4\kappa_c^2\omega_m}{(\Delta+\omega_m)^2+\kappa_c^2}
\end{align}
to accommodate for the ``optical spring effect''~\cite{C.Genes,K.Jahne} described by Eq.(18), the efficient transfer of squeezing can occur, which is described by the master equation of motion for the movable mirror
\begin{align}
\frac{d}{dt}\rho_f&=\frac{\gamma_\textrm{opt}}{2}\vert M\vert e^{i\varphi'}(2f\rho_ff-f^2\rho_f-\rho_ff^2)\nonumber\\
&+\frac{\gamma_\textrm{opt}}{2}\vert M\vert e^{-i\varphi'}(2f^\dag\rho_ff^\dag-f^{\dag2}\rho_f-\rho_ff^{\dag2})\nonumber\\
&+\frac{\gamma_\textrm{opt}}{2}N(2f^\dag\rho_ff-ff^\dag\rho_f-\rho_fff^\dag)\nonumber\\
&+\frac{\gamma_\textrm{opt}}{2}(N+1)(2f\rho_ff^\dag-f^\dag f\rho_f-\rho_ff^\dag f),
\end{align}
where optically-induced damping rate is
\begin{eqnarray}
\gamma_\textrm{opt}=2\frac{\vert g_\textrm{eff}\vert^2}{\kappa_c}\frac{4\omega_m^2}{(\Delta+\omega_m)^2+\kappa_c^2}
\end{eqnarray}
and $\varphi'=\arg\big\{\frac{\bar{a}^2}{[i(\Delta+\omega_m)+\kappa_c]^2}\big\}-\varphi$ is the new squeezing phase factor. It is obvious that the cavity field behaves like the squeezed vacuum reservoir for the movable mirror with the required frequencies of optical reservoir
\begin{align}
\omega_R&=\omega_a+\omega_m/2-\frac{\alpha}{\beta}\sqrt{\frac{2\kappa_cc}{L}}/2,\nonumber\\
\omega_s&\approx\omega_R+\omega_m
\end{align}
due to the negligibility of $\delta$ compared with $\omega_R, \omega_m$. In addition, we can choose appropriate initial phase of the input squeezed vacuum light $\varphi$ or coherent driving light $\bar{a}_{in}$ to make $\varphi'=0$ for simplicity.

Considering the experimental realizable parameters in Refs.~\cite{A.Xuereb} and~\cite{D.Friedrich}, mechanical oscillator's effective mass is $m=100\textrm{ng}$, frequency is $\omega_m=2\pi\times103\textrm{kHz}$, intrinsic damping rate is $\gamma_m=2\pi\times0.025\textrm{Hz}$, cavity's damping rate is $\kappa_c=2\pi\times196\textrm{kHz}$ and the tunable dispersive optomechanical coupling for which $\alpha=0$ can be also achieved. When the input power is $10\textrm{mW}$, the corresponding effective dissipative cavity-mirror coupling strength in this experimentally realized FPI system reaches $2\vert g_\textrm{eff}\vert\approx0.07\kappa_c$, which is well within the weak-coupling regime to validate the adiabatically eliminating approach for the cavity field.

\subsection{Cooling of the movable mirror}
The squeezed-state mechanical mirror has many applications under the conditions of ground-state cooling~\cite{M.D.LaHaye}, therefore cooling down the mechanical oscillator is a vital step toward the practical implementation. In absence of optomechanical coupling the movable mirror is still coupled to the mechanical bath. The mirror is damped at the intrinsic rate $\gamma_m$ which leads to a mean phonon number in thermal equilibrium $n_\textrm{th}$. In presence of the mechanical bath and the optomechanical coupling, the total damping rate $\gamma_\textrm{tot}$ becomes a sum of intrinsic damping rate $\gamma_m$ and optically-induced damping rate $\gamma_\textrm{opt}$
\begin{align}
\gamma_\textrm{tot}=\gamma_m+\gamma_\textrm{opt},
\end{align}
and the steady-state mean phonon number becomes
\begin{align}
n_{st}=(\gamma_mn_\textrm{th}+\gamma_\textrm{opt}N)/(\gamma_m+\gamma_\textrm{opt}).
\end{align}
In fact, for the particular case of no injection of squeezed vacuum noise into the cavity, i.e. $M=N=0$, the final occupation number is $n_{st}=\gamma_mn_\textrm{th}/(\gamma_m+\gamma_\textrm{opt})$. In general, for high-Q mechanical oscillators and efficient laser cooling, it is feasible to take the relation $\gamma_mn_\textrm{th}\ll\gamma_\textrm{opt}$. For example, with the parameters shown in last section, the optically-induced damping rate for the movable mirror becomes $\gamma_\textrm{opt}=2\pi\times320\textrm{Hz}$, which is 4 orders of magnitude higher than the intrinsic damping rate $\gamma_m$. Thus it is possible to achieve ground-state cooling, which is also independent of the ratio $\kappa_c/\omega_m$. These results are consistent with those in Refs.~\cite{F.Elste} and~\cite{A.Xuereb}, which are obtained with the use of the Heisenberg-Langevin approach. The cooling scheme can be physically explained as follows: via utilizing the completely destructive interference of quantum noise, the driven cavity effectively acts as a zero-temperature bath irrespective of the ratio $\kappa_c/\omega_m$, leading the movable mirror to cool down to the ground state.

If we neglect the contribution of the phononic heat bath and feed the squeezed vacuum noise into the cavity, the steady-state mean phonon number is $n_{st}=N=\sinh^2(r)$ calculated from Eq.(25), which coincides with the average input photon number of the reservoir. For example, for the squeezing parameter $r=1$, the phonon number is $n_{st}=1.38$. The movable mirror is very close to the ground state. Based on the elimination of heating process in the dissipative cooling scheme, it also opens the possibility to realize the efficient squeezing transfer from the light field to movable mirror outside the resolved-sideband limit.

\subsection{Squeezing of the movable mirror}
In order to study the squeezing of the movable mirror, we need to evaluate the variances of the generalized quadrature operators
\begin{align}
X=(f+f^\dag)/\sqrt{2}, \hspace{10pt} Y=i(f^\dag-f)/\sqrt{2}.
\end{align}
From the motion equation in Eq.(21), after some calculations we obtain position and momentum fluctuations in a simple form by neglecting the thermal noise
\begin{align}
&\langle X^2\rangle=N+\frac{1}{2}-\vert M\vert=\frac{1}{2}e^{-2r},\nonumber\\
&\langle Y^2\rangle=N+\frac{1}{2}+\vert M\vert=\frac{1}{2}e^{2r}.
\end{align}
Obviously, the position squeezing of the movable mirror occurs and the mirror is in the ideal squeezed state. The squeezing factor of the movable mirror is $r$ and equal to that of the input squeezed noise, which means that the squeezing is perfectly transferred from the light reservoir to the movable mirror in this dissipative optomechanical system. The reason is that since the squeezing for the movable mirror is vulnerable to the heating processes, i.e. the thermal bath and optically-induced heating, the success in the elimination of heating scattering and enhancement of cooling rate without limited by the resolved-sideband regime guarantees the ideal squeezing state for the movable mirror in this dissipative optomechanics.

For the mechanism of transfer of squeezing from light to a membrane based on the resolved-sideband cooling scheme~\cite{K.Jahne}, which is purely dispersive cavity-mirror coupling, ideal squeezed state is only possible under the conditions of the suppressed heating scattering well within the resolved-sideband limit. The squeezing for the mirror starts to degrade outside the resolved-sideband regime because the optically-induced heating process becomes to take into account, which influences the squeezing transfer. In contrast, in this dissipative optomechanics, the movable mirror is in the ideal squeezed state independent upon the ratio of $\kappa_c/\omega_m$ due to the perfect elimination of the optical-induced heating via utilizing destructive interference of quantum noise. The cavity field mimics a ideal squeezed vacuum environment for the movable mirror without requiring the cavity to be in so-called good cavity limit. Moreover, the cooling rate is not limited by the low cavity decay rate, making the squeezed state be robust against the thermal noise. Therefore, the perfect squeezing of the movable mirror close to its ground state can be achieved in the non-resolved-sideband regime. These analytical results for steady-state mean phonon number and the squeezing will be numerically validated in the next section.

\section{Cooling and squeezing beyond the weak-coupling regime}
We have presented the perfect squeezing transfer from the squeezed vacuum light to the movable mirror in the weak-coupling limit in last section, and then we will show the squeezing transfer for the general case of dispersive and dissipative coupling beyond the weak-coupling limit. Since a broadband squeezed vacuum is assumed, the Markovian master equation for the cavity-mirror system obtained via adiabatically eliminating the squeezed vacuum reservoir variables in Eq.(4) is still valid~\cite{H.J.Carmicael}. To proceed, the classical components $\bar{a}$ and $\bar{b}$ are unchanged and now we need the full solutions for Eqs.(10)--(13). For simplicity, we focus on the purely dissipative optomechanics, i.e., $\alpha=0$, which does not influence the physics discussed in the system.

We turn to calculate a close set of motional equations for second moments $\big\{\langle d^2\rangle, \langle d^{\dag2}\rangle, \langle d^\dag d\rangle, \langle d(f+f^\dag)\rangle, \langle d(f-f^\dag)\rangle, \langle d^\dag(f+f^\dag)\rangle, \langle d^\dag(f-f^\dag), \langle (f+f^\dag)^2\rangle, \langle(f-f^\dag)(f+f^\dag)\rangle, \langle(f-f^\dag)^2\rangle\big\}$, from which we will obtain the steady-state mean phonon number and squeezing for the movable mirror. These motional equations are presented in the Appendix, and there we obtain the steady-state mechanical occupation number in Eq.(A.7). For the moderately strong input driving fields and under the conditions of $\Delta=\omega_m/2$, we expand the result up to first order in the $\vert g_\textrm{eff}\vert^2/\kappa_c^2$, which becomes
\begin{align}
n_{st}=N+(1+2N)\vert g_\textrm{eff}\vert^2/\kappa_c^2.
\end{align}
The term proportional to $\vert g_\textrm{eff}\vert^2/\kappa_c^2$ corresponds to optical-induced heating for the movable mirror, which is resulted from the non elimination of photonic excitation as compared with the weak-coupling regime. It indicates that photonic excitation precludes the complete destructive interference of the quantum noise appeared in the weak-coupling regime. When we replace the input squeezed reservoir by the vacuum, i.e, $N=0$, the result coincides with the expression in Ref.~\cite{A.Xuereb} by neglecting the intrinsic damping rate. On the other hand, the optimal $\langle f^2\rangle$ is related to well-chosen $\Delta_s$ to accommodate for cavity induced energy shift for movable mirror, and we can numerically find the appropriate detuning $\Delta_s$ around $\omega_m$ to obtain the optimum squeezing state for the movable mirror.
\begin{figure}[hbt]
\centering\includegraphics[width=7.5cm,keepaspectratio,clip]{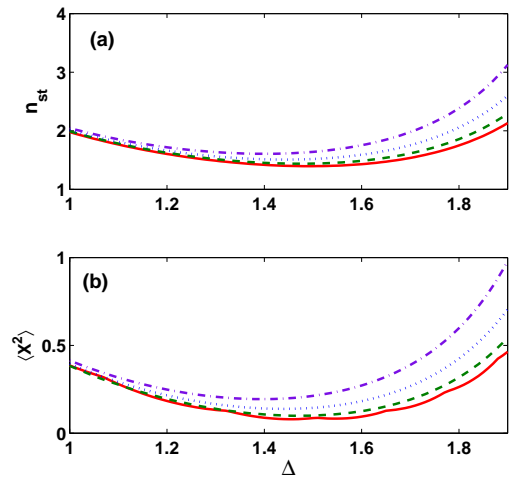}
\caption{ (Color online) The steady-state mean phonon number $n_{st}$ in (a) and the position squeezing $\langle X^2\rangle$ in (b) as functions of the detuning $\Delta$ with the different effective coupling strengths between the cavity field and movable mirror, with the parameters (in units of $\kappa_c$) $\omega_m=3\kappa_c$, $\Delta_s\approx\omega_m$, $r=1$, and $\alpha=0$. The effective coupling strengths are taken as $G_\textrm{eff}/\kappa_c$={0.1 (red solid line), 0.2 (green dashed line), 0.3 (blue dotted line), 0.4 (purple dash-dotted line)} respectively and the optimal squeezing is obtained via carefully tuning $\Delta_s$ around $\omega_m$.}
\end{figure}

We numerically calculate the steady-state mean phonon number and the squeezing for the position operator with arbitrary effective coupling between cavity mode and movable mirror characterized by $G_\textrm{eff}=2\vert g_\textrm{eff}\vert$, and the numerical results are demonstrated in Fig.2. The minimum phonon number and the optimal squeezing are achieved at $\Delta=\omega_m/2$ clearly, which coincides with result in Eq.(19). In special, in the weakly coupling regime, for example, $G_\textrm{eff}/\kappa_c=0.1$ indicated by the red solid curve, the numerical results $n_{st}=1.395$ and $\langle X^2\rangle=0.08$ around $\Delta=\omega_m/2$, agree with the corresponding analytical results which are 1.38 and 0.068 obtained from Eqs.(25) and (27). Including the higher-order correction, steady-state mean phonon number 1.39 is better agreement with the numerical result, which is calculated from Eq.(28).

In addition, for the moderate coupling strength, the incomplete destructive quantum interference hinders the optimal cooling for the movable mirror because of the existence of higher-order optical-induced heating in $\vert g_\textrm{eff}\vert^2/\kappa_c^2$. Simultaneously, $\vert\langle f^2\rangle\vert$ can not be larger than $M$. The resulting relation
\begin{align}
\sqrt{n_{st}(n_{st}+1)}>\vert\langle f^2\rangle\vert
\end{align}
is fulfilled, which means that the mirror deviates from the ideal squeezed state with the increased coupling strength. However, the squeezed state for the movable mirror can still occur beyond the weak-coupling regime indicated in Fig.2, in which the curves demonstrate the ability of dissipative optomechanical system in producing the squeezing for the position operator around its motional ground state.

\section{Conclusion}
In conclusion, we present an optomechanical system consisted of an effective FPI with one movable ideal end mirror, which is capable of generating mechanical squeezing via engineering reservoir. Via feeding a much weaker broadband squeezed vacuum light accompanying the coherent driving laser field into the cavity, the cavity field is coupled to the movable mirror through both the tunable dispersive and dissipative interactions. The motion equation for the cavity-mirror system is derived by following the general reservoir theory based on the density operator in which the reservoir variables are adiabatically eliminated by using the reduced density operator for the system in the interaction picture. When the mirror is weakly coupled to the cavity mode, the driven cavity can effectively perform as a squeezed vacuum reservoir for the movable mirror under the conditions of laser cooling to its ground motional state, where the optically-induced heating scattering is perfectly eliminated via the complete destructive interference of quantum noise. Thus, the perfect transfer of squeezing from the light to the movable mirror occurs, which is irrespective of the ratio between the cavity damping rate and the mechanical frequency. When the mirror is coupled to the cavity field beyond the weak-coupling regime, the photonic excitation can preclude the complete destructive interference of quantum noise, leading to the mirror deviation from the ideal squeezed state. However, the dissipative optomechanics can still produce the squeezed state for the mirror for the moderately coupling strength.

{\bf{Acknowledgements:}}
This work is supported by the National Natural Science Foundation of China (Grant Nos. 11074087 and 61275123), the Nature Science Foundation of Wuhan City (Grant No. 201150530149), the National Basic Research Program of China (Grant No. 2012CB921602), and the Key Laboratory of Advanced Micro-Structure of Tongji University.

\appendix*
\section{}
We derive the motional equations for the second-order moments of the cavity and the mirror variables from the Eqs.(10)-(13), which are written into a vector form as
\begin{align}
\vec{X}=&\Big(\langle d^2\rangle, \langle d^{\dag2}\rangle, \langle d^\dag d\rangle, \langle df_+\rangle, \langle df_-\rangle, \langle d^\dag f_+\rangle,\nonumber\\& \langle d^\dag f_-\rangle, \langle f_+^2\rangle, \langle f_-f_+\rangle, \langle f_-^2\rangle\Big)^T,
\end{align}
where $f_+=f+f^\dag, f_-=f-f^\dag$ and $T$ denotes the transpose of the vector. The second moments obey the equation
\begin{align}
\frac{d}{dt}\vec{X}=\underline{A}\vec{X}+B^{(+)}e^{i2\Delta_st}+B^{(-)}e^{-i2\Delta_st}+B^{(0)},
\end{align}
where the coefficient matrix $\underline{A}$ is
\begin{widetext}
\begin{align}
\underline{A}=\left(
                \begin{array}{cccccccccc}
                  2(i\Delta-\kappa_c) & 0 & 0 & \xi & 0 & 0 & 0 & 0 & 0 & 0 \\
                  0 & -2(i\Delta+\kappa_c) & 0 & 0 & 0 & \xi^* & 0 & 0 & 0 & 0 \\
                  0 & 0 & -2\kappa_c & \xi^*/2 & 0 & \xi/2 & 0 & 0 & 0 & 0 \\
                  0 & 0 & 0 & i\Delta-\kappa_c & -i\omega_m & 0 & 0 & \xi/2 & 0 & 0 \\
                  \chi^* & 0 & -\chi & -i\omega_m & i\Delta-\kappa_c & 0 & 0 & 0 & \xi/2 & 0 \\
                  0 & 0 & 0 & 0 & 0 & -(i\Delta+\kappa_c) & -i\omega_m & \xi^*/2 & 0 & 0 \\
                  0 & -\chi & \chi^* & 0 & 0 & -i\omega_m & -(i\Delta+\kappa_c) & 0 & \xi^*/2 & 0 \\
                  0 & 0 & 0 & 0 & 0 & 0 & 0 & 0 & -i2\omega_m & 0 \\
                  0 & 0 & 0 & \chi^* & 0 & -\chi & 0 & -i\omega_m & 0 & -i\omega_m \\
                  0 & 0 & 0 & 0 & 2\chi^* & 0 & -2\chi & 0 & -i2\omega_m & 0 \\
                \end{array}
              \right)
\end{align}
\end{widetext}
with $\chi=2g_0\beta\sqrt{\frac{L}{c}}\bar{a}_{in}, \zeta=4g_\textrm{eff}, \xi=-\chi-\zeta$, and the nonhomogeneous terms are
\begin{align}
B^{(+)}&=\big(0, \hspace{2pt}2\kappa_cM^*, \hspace{2pt}0, \hspace{2pt}0, \hspace{2pt}0, \hspace{2pt}0, \hspace{2pt}-\zeta M^*, \hspace{2pt}0, \hspace{2pt}0, \hspace{2pt}\frac{\zeta^2M^*}{2\kappa_c}\big)^T,\nonumber\\
B^{(-)}&=\big(2\kappa_cM, \hspace{2pt}0, \hspace{2pt}0, \hspace{2pt}0, \hspace{2pt}\zeta^*M, \hspace{2pt}0, \hspace{2pt}0, \hspace{2pt}0, \hspace{2pt}0, \hspace{2pt}\frac{\zeta^{*2}M}{2\kappa_c}\big)^T,\nonumber\\
B^{(0)}&=\big(0, \hspace{2pt}0, \hspace{2pt}2\kappa_cN, \hspace{2pt}0, \hspace{2pt}-\zeta N, \hspace{2pt}0, \hspace{2pt}\chi^*+\zeta^*(N+1), \hspace{2pt}i2\omega_m, \nonumber\\&\hspace{2pt}0, \hspace{2pt}i2\omega_m-(2N+1)\frac{\vert\zeta\vert^{2}}{2\kappa_c}\big)^T.
\end{align}
We expand the time-dependent $\vec{X}$ into a sum of the slowly varying components composed of $\vec{X}^{(0)}, \vec{X}^{(+)}, \vec{X}^{(-)}$ with the oscillating frequencies $0$, $2\Delta_s$, $-2\Delta_s$,
\begin{align}
\vec{X}=\vec{X}^{(0)}+\vec{X}^{(+)}e^{i2\Delta_st}+\vec{X}^{(-)}e^{-i2\Delta_st}.
\end{align}
Thus the steady-state solutions for the Eq.(A.2) are given as
\begin{align}
&\vec{X}^{(0)}=-\underline{A}^{-1}B^{(0)},\nonumber\\
&\vec{X}^{(+)}=(i2\Delta_s-\underline{A})^{-1}B^{(+)},\nonumber\\
&\vec{X}^{(-)}=-(i2\Delta_s+\underline{A})^{-1}B^{(-)},
\end{align}
from which we obtain the mean phonon number $\langle f^\dag f\rangle$ and $\langle f^2\rangle$.
The steady-state mean phonon number $\langle f^\dag f\rangle$ takes the form
\begin{align}
\langle f^\dag f\rangle&=N+\frac{1+2N}{4}\nonumber\\&\times\bigg\{\frac{\omega_m(\Delta^2-\kappa_c^2)(\omega_m-2\Delta)-\Delta^2(\Delta^2+\kappa_c^2)}{\omega_m\Delta(\Delta^2-\kappa_c^2)}\nonumber\\&
+\frac{(\Delta^2+\kappa_c^2)^2}{\Delta^2-\kappa_c^2}\Big[\frac{-\Delta(\Delta^2+\kappa_c^2)}{\chi^2\Delta(\Delta^2-3\kappa_c^2)+\omega_m(\Delta^2+\kappa_c^2)^2}\nonumber\\&
+\frac{\chi^2\Delta/2-(\Delta^2+\kappa_c^2)\omega_m}{\Delta(\Delta^2+\kappa_c^2)(2\Delta^2-2\kappa_c^2-\omega_m^2)+\chi^2\Delta^2\omega_m}\Big]\bigg\}.
\end{align}
Under the conditions of the optimal detuning $\Delta=\omega_m/2$, we expand $\langle f^\dag f\rangle$ up to the order in $\vert g_\textrm{eff}\vert^2/\kappa_c^2$ and obtain
\begin{align}
\langle f^\dag f\rangle=N+(1+2N)\vert g_\textrm{eff}\vert^2/\kappa_c^2.
\end{align}
On the other hand, the optimal $\langle f^2\rangle$ is related to the well-chosen detuning $\Delta_s$ to accommodate for the cavity induced energy shift for the movable mirror, and here we can numerically find out the $\Delta_s$ to obtain the optimum squeezing, which is discussed in the main text.


\begin{references}
\bibitem{I.WalsonRae} I. Walson-Rae, N. Nooshi, W. Zwerger, and T. J. Kippenberg, Phys. Rev. Lett. {\bf 99}, 093901 (2007).
\bibitem{F.Marquardt1} F. Marquardt, J. P. Chen, A. A. Clerk, and S. M. Girvin, Phys. Rev. Lett. {\bf 99}, 093902 (2007).
\bibitem{T.J.Kippenberg} T. J. Kippenberg, and K. J. Vahala, Science {\bf 321}, 1172 (2008).
\bibitem{F.Marquardt2} F. Marquardt, and S. M. Girvin, Physics {\bf 2}, 40 (2009).
\bibitem{V.Fiore} V. Fiore, Y. Yang, M. C. Kuzyk, R. Barbour, L. Tian, and Hailin Wang, Phys. Rev. Lett. {\bf 107}, 133601 (2011).
\bibitem{K.Stannigel} K. Stannigel, P. Rabl, A. S. Sorensen, M. D. Lukin, and P. Zoller, Phys. Rev. A {\bf 84}, 042341 (2011).
\bibitem{haoxiong} H. Xiong, L.-G. Si, X.-Y. L\"u, X. X. Yang, and Y. Wu, Opt. Lett. {\bf 38}, 353-355 (2013).
\bibitem{P.Verlot}P. Verlot, A. Tavernarakis, T. Briant, P.-F. Cohadon, and A. Heidmann, Phys. Rev. Lett. {\bf 104}, 133602 (2010).
\bibitem{E.Gavartin} E. Gavartin, P. Verlot, and T. J. Kippenberg, Nature Nanotechnology {\bf 7}, 509-514 (2012).
\bibitem{JianQiZhang} J.-Q. Zhang, Y. Li, M. Feng, and Y. Xu, Phys. Rev. A {\bf 86}, 053806 (2012).
\bibitem{I.Pikovski} I. Pikovski, M. R. Vanner, M. Aspelmeyer, M. S. Kim, and C. Brukner, Nat. Phys. {\bf 8} 393 (2012).
\bibitem{A.Schliesser} A. Schliesser, O. Arcizet, R. Rivi\`ere, G. Anetsberger, and T. J. Kippenberg, Nature Physics {\bf 5}, 509-514 (2009).
\bibitem{L.Mazzola} L. Mazzola, and M. Paternostro, Scientific Reports {\bf 1}, 199 (2011).
\bibitem{DWC.Brooks} D. W. C. Brooks, T. Botter, S. Schreppler, T. P. Purdy, N. Brahms, and D. M. Stamper-Kurn, Nature {\bf 488}, 476¨C480, (2012).
\bibitem{S.Rips} S. Rips, M. Kiffner, I. Wilson-Rae, and M J Hartmann, New J. Phys. {\bf 14}, 023042 (2012).
\bibitem{C.M.Caves} C. M. Caves, K. S. Thorne, R. W. P. Drever, V. D. Sandberg, and M. Zimmermann, Rev. Mod. Phys. {\bf 52}, 341 (1980).
\bibitem{A.Abramovici} A. Abramovici \emph{et al.}, Science {\bf 256}, 325 (1992).
\bibitem{B.C.Barish} B. C. Barish, and R. Weiss, Phys. Today {\bf 52}, 44 (1999).
\bibitem{K.Jahne} K. J\"ahne, C. Genes, K. Hammerer, M. Wallquist, E. S. Polzik, and P. Zoller, Phys. Rev. A {\bf 79}, 063819 (2009).
\bibitem{A.Mari} A. Mari, and J. Eisert, Phys. Rev. Lett. {\bf 103}, 213603 (2009).
\bibitem{HuatangTan} H. T. Tan, G. X. Li, and P. Meystre, Phys. Rev. A {\bf 87}, 033829 (2013).
\bibitem{J.Q.Liao} J. Q. Liao, and C. K. Law, Phys. Rev. A {\bf 83}, 033820 (2011).
\bibitem{Sumei} Sumei Huang, and G. S. Agarwal, Phys. Rev. A {\bf 79}, 013821 (2006).
\bibitem{H.Seok} H. Seok, L. F. Buchmann, S. Singh, S. K. Steinke, and P. Meystre, Phys. Rev. A {\bf 85}, 033822 (2012).
\bibitem{F.Elste} F. Elste, S. M. Girvin, and A. A. Clerk, Phys. Rev. Lett. {\bf 102}, 207209 (2009).
\bibitem{A.Xuereb} A. Xuereb, R. Schnabel, and K. Hammerer, Phys. Rev. Lett. {\bf 107}, 213604 (2011).
\bibitem{K.Yamamoto} K. Yamamoto, D. Friedrich, T. Westphal, S. Gobler, K. Danzmann, K. Somiya, S. L. Danilishin, and R. Schnabel, Phys. Rev. A {\bf 81}, 033849 (2010).
\bibitem{D.Friedrich} D. Friedrich, H. Kaufer, T. Westphal, K. Yamamoto, A. Sawadsky, F. Y. Khalili, S. L. Danilishin, S. Gobler, K. Danzmann and R. Schnabel, New J. Phys. {\bf 13}, 093017 (2011).
\bibitem{T.Weiss} T. Weiss, C. Bruder, and A. Nunnenkamp, arXiv:1211.7029 [quant-ph].
\bibitem{C.W.Gardiner} C. W. Gardiner, Phys. Rev. Lett. {\bf 56}, 1917 (1986).
\bibitem{H.J.Carmicael} H. J. Carmichael, A. S. Lane, and D. F. Walls, Phys. Rev. Lett. {\bf 58}, 2539--2542 (1987).
\bibitem{M.O.Scully} M. O. Scully, and M. S. Zubairy, \emph{Quantum optics} (Cambridge university press, 1997).
\bibitem{A.Nunnenkamp} A. Nunnenkamp, K. Borkje, and S. M. Girvin, Phys. Rev. Lett. {\bf 107}, 063602 (2011).
\bibitem{W.J.Gu} W. J. Gu, and G. X. Li, Phys. Rev. A {\bf 87}, 025804 (2013).
\bibitem{C.Genes} C. Genes, D. Vitali, P. Tombesi, S. Gigan, and M. Aspelmeyer, Phys. Rev. A {\bf 77}, 033804 (2008).
\bibitem{M.D.LaHaye} M. D. LaHaye, O. Buu, B. Camarota, and K. C. Schwab, Science {\bf 304}, 74 (2004).
\end{references}
\end{document}